MIT Lincoln Laboratory

# Simulated Automatic Exposure Notification (SimAEN): Exploring the Effects of Interventions on the Spread of COVID

Ted Londner, Jonathan Saunders, Dieter W. Schuldt, Dr. Bill Streilein

# Contents

# Introduction

Speed is critical to the success of a response to an epidemic. During an uncontrolled outbreak the number of cases grows exponentially, so removing infected individuals from the population will have tremendous payoffs in terms of reducing the overall number of infections.

Mitigation strategies that remove infectious individuals from the greater population have to balance their efficacy with the economic effects associated with quarantine and have to contend with the limited resources available to the public health authorities. Prior strategies have relied on testing and contact tracing to find individuals before they become infectious and in order to limit their interactions with others until after their infectious period has passed.

Manual contact tracing (MCT) is a public health intervention where individuals testing positive are interviewed to identify other members of the community who they may have come into contact with. These interviews can take a significant amount of time that has to be tallied in the overall accounting of the outbreak cost.

The concept of contact tracing has been expanded recently into "Automatic Electronic Notification" (AEN) or "Electronic Contact Tracing" whereby individual's cellphones can be used as sensor platforms to log close contacts and notify them in the event that one of their close contacts tests positive. The intention is that this notification will prompt the person to be tested and then restrict their interactions with others until their status is determined.

In this paper we describe our efforts to investigate the effectiveness of contact tracing interventions on controlling an outbreak. This is accomplished by creating a model of disease spread and then observing the impact that simulated tracing and testing have on the number of infected individuals. Model parameters are explored to identify critical transition points where interventions become effective. We estimate the benefits as well as costs in order to offer insight to public health officials as they select courses of action.

## Prior Modeling Efforts

The prediction and estimation of disease spread has been performed using many different methods. In general, these methods land in three broad categories: compartmental models, data-driven models, and agent-based models.

### Compartmental models

Compartmental models use a series of interdependent ordinary differential equations (ODEs) to describe the mean number of individuals in each of several categories. The total population is broken down into several subpopulations (susceptible, exposed, infectious, and recovered among others) and the fraction of the population in each of the subpopulations changes over time according to the prescribed ODEs (Kermack & McKendrick, 1927). These models are typically described using a convenient shorthand (SIR, SEIR, etc.) indicating the subpopulations being considered. Owing to their relatively easy evaluation and easily understandable results, these models have gained wide adoption in the epidemiological community. To gain this tractability the model assumes that the entire population is homogeneously mixed. However, this assumption and other aspects of their formulation limits their ability to predict the real-world operational changes that occur when resources are limited and interactions are conditionally probabilistic. Small world models, such as those by Strogatz and Watts

(Watts & Strogatz, 1998), show that mixing heterogeneity is omnipresent in human interaction and greatly affects the propagation of disease.

### Data-driven models

The idea behind data-driven modeling is that given sufficient data about a population you can discover a relationship among the inputs that minimizes the error signal between the model and some set of real world outputs. This can take the form of neural networks (Wieczorek, Siłka, & Woźniak, 2020), fuzzy logic systems (Traulsen & Krieter, 2012), or regressions (Werneck, et al., 2002) which all use prior data to train the weights in the systems of equations that constitute the model. One problem with these models is that they can be opaque, meaning their internal workings are difficult to interpret and their outputs hard to justify.

This sort of model does not consider the population dynamics explicitly, instead focusing on measurable data and assuming that there is an underlying sensible reality generating them. By creating a system with enough freedom, the model is able to work as a surrogate for this reality. However, since they are only a product of the data expert knowledge might not be considered.

### Agent-based models

Agent-based models attempt to limit abstraction, instead depicting every member of the susceptible population and adjusting their condition based on the progression of the disease according to a set of rules.

On the most abstract side of agent-based modeling are grid based simulations known as cellular automata. This framework is most famously implemented in Conway's game of life (Adamatzky, 2010) but also been applied to disease dynamics (White, del Rey, & Sanchez, 2007). These models treat the agents as nodes of a regular grid or irregular network (Gagliardi & Alves, 2010) and update the state of these agents based on the state of their neighbors.

Less abstract agent-based models account for spatial variation in interactions. These models create agents that conform to 'schedules' as they move through a virtual landscape. The population of agents can mirror reality in terms of demographics, behavior, and spatial distribution. Underlying these models is the assumption that if reality is intricately modeled then the model will behave in the same way as reality.

All of this verisimilitude comes at a cost. The amount of computation required to evaluate the model goes up as a function of the number of agents and memory requirements expand as the list of agent parameters (e.g., features) grows. For these reasons, these models are reasonably recent developments, only just now being fully realizable thanks to modern computing capabilities.

## Model Definition

The SimAEN model definition is based on the ODD protocol as specified in (Grimm, et al., 2010). This formulation establishes a standard way of communicating an agent-based model in a manner that allows for re-creation by other interested parties. The agents in this description represent people and we will refer to them as agents, people, or individuals interchangeably.

## Overview

### Purpose

AEN is a new technology and there is need to understand its effectiveness. The SimAEN agent-based simulation was developed to explore and understand how AEN and its internal configuration and components affect disease spread. In addition to the independent effects of AEN we also endeavor to show the effects of manual contact tracing, widespread testing, and the results of deploying these strategies in combination with each other.

### Entities, state variables, and scales

Agents in the SimAEN model represent individuals whose interactions are guided by a collection of probabilities. These individuals operate within the context of a "world" which also has a collection of probabilities and parameters assigned to various aspects of it (e.g, the duration of each phase of the disease, the probability of a call from public health being answered by an individual, etc.). Agent-based models are naturally suited to object-oriented programming methods, so both individuals and worlds can be thought of as objects -- though for this simulation only a single world exists at any one time.

Further information on the parameters of the individual and world objects can be found in the appendix. These parameters describe the various states that individuals can be in during the simulation.

The world in which individuals exist advances on a discrete schedule, where state changes occur once per day. This timeframe (e.g., once per day) was chosen because it captures the phenomena relevant to disease transmission, while also being long enough that it does not take an unacceptable amount of computation to determine results. In addition, longer timeframes, such as (e.g., once per week) would not permit the granularity associated with testing events or other characteristics of interest.

The individual is the smallest agent level consideration. This model does not directly consider family units or workplace structures. The type of transmission events that take place in these settings are accounted for by modeling the events using a log-normal distribution. This distribution features a long tail, meaning that there is a relatively high probability of an event occurring where a large number of individuals contract the disease (e.g., in the tail of the distribution).

### Process overview and scheduling

The simulation advances one day at a time, during which each of the individuals in the simulation updates its status, potentially gets tested, is processed by public health, and changes its behavior.

Each day is broken down into a series of events where various aspects of the simulation are performed, These events always occur in the specified order, though this order was chosen arbitrarily.

The first event to be processed is transmission. During this event each individual is evaluated to see if it produced additional infections. For each new infection produced, new individuals are instantiated. Some number of uninfected individuals are also produced, based on the false detection rate (FDR) and the probability of transmission associated with the individual.

Following the transmission event comes the testing event during which all individuals are checked to see if they get a test based on a probability conditioned on the agent's traits. If a test is performed then a countdown is started, simulating the delay that occurs between test and results. It is assumed that the testing capacity is larger than the number of cases and thus the delay between test and result is not a function of the number of people who are being tested.

The next event is automatic tracing, where notifications are sent out by individuals who have tested positive and have probabilistically decided to upload their keys to the AEN application. Receiving a notification from the AEN system will change the individual's probability of getting tested and also altering their behavior.

In the final event, public health performs manual contact tracing. This step involves contacting a person who has tested positive to identify individuals they may have come in contact with and have potentially infected.

At the beginning and end of each day various bookkeeping tasks are executed. This includes removing individuals from the simulation if they have existed for four disease generations, since there is no way for these individuals to pertain to the current state and propagation of the simulation. This saves on memory and future computational cost.

## Design concepts

### Basic principles

This SimAEN model is based on the transmission and public health response associated with COVID-19. This means that transmission occurs through close contact of individuals, not through contact with a previously exposed surface, consumption of contaminated food or drink, or other methods of disease transfer. This affects the number of people that can be expected to become infected by an individual in any particular transmission event and the likelihood that a person would be able to identify the person that infected them or who they may have infected.

One important assumption of our model is that the disease only ever infects a small portion of the population. In the standard SEIR model the rate of change of the susceptible population (i.e., $dS$) is a function of the infected population as a proportion of the overall population. However, as long as this fraction is small we can treat the susceptible population as a constant; a pool of individuals that we can always draw from. As such, we do not model the greater population of people who have not been directly affected by the disease. Instead, individuals are created at the time that they are needed and then disposed of once they are no longer integral to the simulation. The downside of this method is that it allows the number of people affected to grow without bound over long time scales. However, it saves significant computation by not simulating all of the people who are not impacted (which, per our assumption, is most of them). This also allows us to ignore the spatial aspects of individuals since transmission events do not have to occur at a given intersection of modeled individuals' journeys.

The nature of COVID-19 also determines ranges of potential parameters for stages and duration of the disease. Infections are assumed to last 14 days from initial exposure to recovery. In SimAEN individuals who are infected progress through the following stages of the disease: EXPOSED, PRESYMPTOMATIC, A/SYMPTOMATIC, and RECOVERED.  The SimAEN model does not account for the difference between recovery and death as both of these outcomes remove the individual from the system and do not have an impact on the methods of mitigation employed by public health.

Testing capability is also modeled based on what has been seen in the COVID-19 pandemic. Evidence from the current testing regimen suggests that there are very few false positive results. There is, however, a relatively high rate of false negatives. These rates are dependent on what stage of disease the individual is in at the time that the test is performed.

### Emergence

First, a note on behavioral transition. One of the most famous examples is in random graphs, where it can be shown (Erdős & Rényi, 1960) that as the number of edges in a random graph is increased there is point where the graph rapidly goes from many small components to a state where the vast majority of nodes belong to a single giant component. Many other systems exhibit a similar fundamental change as one (or some collection of) properties is varied.

In simulating AEN, we are interested in finding when model behavior changes – or transitions – as a result of public health interventions. Due to delays caused by limited numbers of contact tracers there is potentially a significant delay between an individual testing positive and public health identifying their contacts. During this delay the disease will continue to be spread. There are also critical transition values for the level of application adoption: below a certain level of adoption, it is unlikely that both the spreader and the newly infected will be running the contact tracing application. Finding where these transitions occur in AEN is important to properly fund public health and to appropriately market the application.

### Adaptation

The SimAEN model assumes that there are four distinct levels of interaction: NORMAL, MINIMAL RESTRICTION, MODERATE RESTRICTION, and MAXIMAL RESTRICTION. Agents with more restrictive behaviors have fewer close contacts, leading to lower levels of disease spread. Each day agents are checked to see if they transition to a more restrictive level. The probability of these transitions is conditioned on several traits (such as receiving a positive test or being contacted by public health) that may also change during the course of the simulation.

Agents will also probabilistically don masks, conditioned on their level of interaction. It is assumed that as agents isolate themselves they are also more likely to take other precautions. Once an agent starts wearing a mask it will not stop wearing a mask. The effects of disease spread are also considered reciprocally, with spread being least likely when both the infected agent and the close contact are wearing masks.

The SimAEN model does not adapt agent behavior based on the progression of time or the prevalence of the disease. That is, agents do not respond to high disease rates by altering their levels of interaction or deciding to wear masks. Changes resulting from public health messaging such as deciding to promote mask wearing are also not modeled.

### Objectives

Since the adaptation in SimAEN is purely driven by probabilities, there is no objective that is trying to be optimized. In the abstract sense the agents are trying to minimize the amount of exposure to others, but this is projected on them by the selection of probabilities for behavior transition.

### Sensing

Agents are aware only of themselves with the exception that they are able to identify some fraction of the people who they have interacted with for the purposes of contact tracing by public health. Agents are able to identify if they are symptomatic, as this is a trait that will affect their probability of getting tested or changing their behavior.

### Interaction

Agents do not interact directly with each other. When an individual transmits a disease it creates new individuals who may or may not be infected. During automated and manual contact tracing individuals interact through an intermediary, either the AEN application or the public health system.

### Stochasticity

This model is driven significantly by stochasticity associated with the initial parameter settings. These probabilities are outlined in the appendix, but a brief overview is given here.

Several agent traits have to be set during agent creation. First, the agent is probabilistically set to be wearing a mask. The mask wearing state of the new agent is combined with the mask state of the infecting agent to determine whether the new agent is set as infected or uninfected. Finally, the new agent has some probability that they had previously downloaded and are running the AEN application.

During transmission events the number of individuals that are created is probabilistic, conditioned on the behavior state of the transmitting individual. If the transmitting individual is running the app there is also the probability that they will generate some number of uninfected agents who were not close enough to have been infected but were identified by the AEN application. This false detection probability models inaccuracies Bluetooth detection capability of digital contact tracing. If the person transmitting the disease and the person being infected are both running the app there is some probability that the app on either end of the transmission will detect the signal.

Whether an agent gets a test is probabilistic, conditioned on whether they have been contacted by public health, received an notification from AEN, tested positive, or are feeling symptomatic.

The probability of a True Positive test is based on the stage of the disease the agent is in at the time of the test. As noted in the assumptions, there are no False Positive test results.

There is a probability that after receiving a positive test an individual will contact public health for the purposes of contact tracing. During contact tracing there is a probability that any given call will successfully reach the agent. If the return call from public health is successful then there is a probability that the traced person infected will be identified.

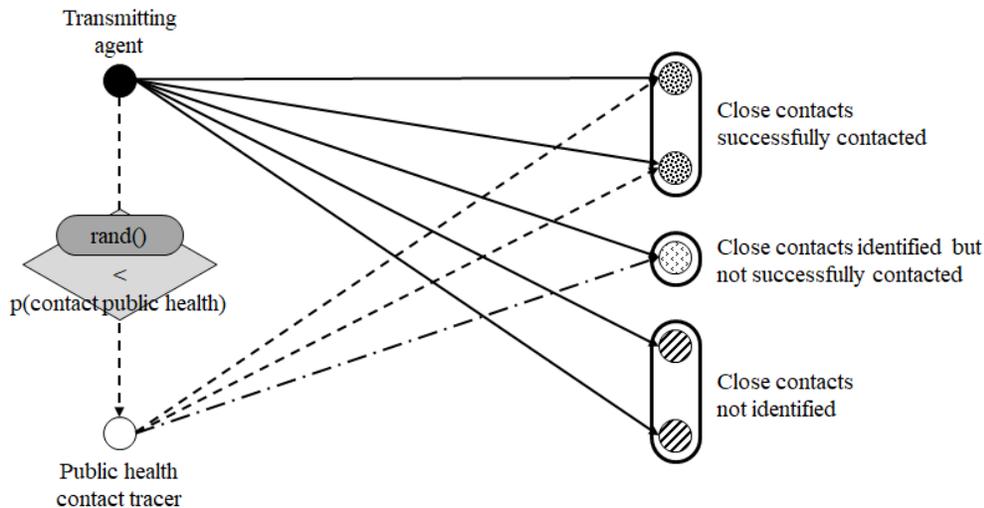

*Figure 1-Diagram of agent generation showing the conditions that agents can obtain based on whether they are identified by the generator*

A positive test also comes with the probability that the individual will upload their key to the AEN server. Uploading a key initiates the AEN, alerting all of their close contacts who are also running the app.

The disease progresses along a trajectory that is different for each individual. This is manifest by randomly selecting the latent and incubation periods on a per person basis. Both of these are sampled from a normal distribution.

Finally, there are probabilities associated with an individual changing their interaction behavior, conditioned on the traits of the individual. There are four interaction states (described above) and the probability of transition is only in a more restrictive direction, except when an individual receives a negative test, in which case they will probabilistically transition to a behavior state based on the distribution of states currently occupied across the simulation.

## Collectives

There are no collectives, such as families or work groups, included in this model. All individuals are treated based only on their own traits and do not have and probabilities or parameters based on their generator or the agents they are generated alongside.

## Observation

Each agent keeps track of every transition that it makes while being simulated. Tracking this information supports forensic analysis of the simulation and how it progress. Examples of information that agents track include all of their traits (such as whether they have been tested) along with the days (since simulation start) on which those traits changed. They also keep track of all generation events including the status of all of the people involved in that event. Agents also track things for which they do not have direct knowledge. For example, agents keep track of how many times they have been unsuccessfully called by public health. This information is not used by the agent but supports deeper understanding of the simulation AEN process.

## Details

### Initialization

The simulation begins with 20 infected individuals. This was chosen because it is a small enough number that it will not overwhelm the steady state yet is large enough that the disease will be able to take hold and propagate. It is assumed that there has been some low level of the disease circulating in the population prior to the exponential grown segment of the spread, and so these 20 individuals are initiated at a random point in the disease progression. This is accomplished by starting them with a "day in system" variable set to a random value drawn from a uniform distribution over the 14 days of the disease lifespan. Each run starts with a new random specification of these 20 agents. These individuals are also assumed to not have been tested, since we are starting from a time before widespread testing is available and AEN is in use.

### Input data

All information necessary for progression of the model is contained within the simulation itself. The only outside input is through the selection of the parameters that drive the system. Once the model is started it carries forward without any additional input.

### Submodels

The SimAEN model includes submodels for testing, behavior, and public health interventions (ACT and MCT).

The testing submodel assumes that individuals will seek out a test contingent on their traits. Each day a random draw occurs for each individual to determine if they will get a test performed. When an individual gets a test it notes its condition on the day when the test is performed and sets a random counter of days until a test result is obtained. When the counter expires, a random draw is made to determine the result of the test, conditioned on the state the agent was in when they were tested. A positive test will further prompt a random determination of whether they will upload their key to the AEN system and/or call public health. Contacting public health will add them to a list of index cases.

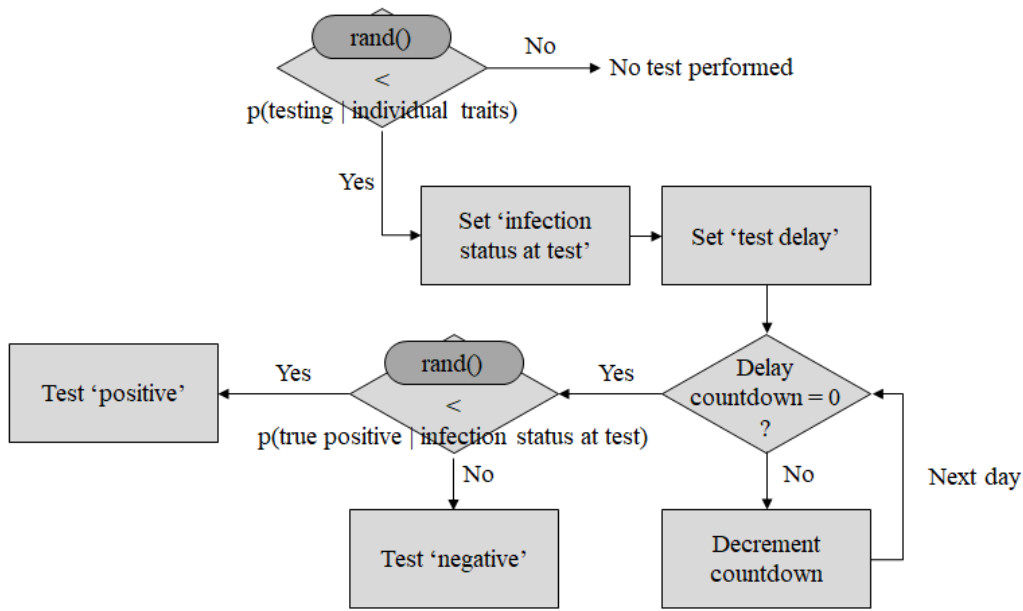

*Figure 2-Diagram of testing methodology as implemented in SimAEN*

Behavioral changes of the individual are based on their traits. All agents start out in the normal – or base state – where they are interacting with society in the way that they would have were they not infected. Each day a random draw is made to determine if the individual will transition to a more restrictive interaction level. It is assumed that an individual will only move to more restrictive behavioral states, with the exception of individuals who are not symptomatic who receive a negative test result. While people in the real world may have day-to-day-variation in their level of interaction we assume that the probabilistic nature of transmission employed in this model is sufficient to describe this and that variation will not significantly affect the measures of interest being studied. There is no prohibition on an agent moving through multiple levels of restriction on a single day assuming that they pass multiple probabilistic draws.

The public health intervention submodels are broken into one each for AEN and MCT.

### AEN

The AEN submodel is relatively straight forward. Individuals receiving a positive test have a random probability of uploading their keys to the AEN system. Uploading a key triggers the AEN system to transmit an alert to all close contacts of the uploader who were running the AEN application and received a beacon message during the transmission even. All of these agents will update their object variables to note that a notification was received.

### MCT

Manual contact tracing (MCT) encompasses both contact tracing and notification of identified contacts. It is assumed that there is a limited number of contact tracers and that each works some period of time each day. Each call they make takes some amount of time, which is based on whether it is a call for the purposes of contact tracing or just as a notification. There is also some time associated with missed calls. Our manual contact tracing submodel revolves around the "call list" which is simply a list of all of the people that need to be called and the "index list" which is the list of individuals who were added to the call list because of a positive list. During the manual contact tracing portion of the main (e.g. daily) simulation evaluation loop calls go out to people in the order they were placed on the call list. When a call occurs the appropriate amount of time is subtracted from the available call time:

$$available\ call\ time = \#\ contact\ tracers \cdot \#\ work\ hours/day$$

If the call being made is to an individual on the index list then a contact trace is performed. If any individuals are identified during the trace they are added to the contact list. Individuals who are not successfully called are added to the back of the contact list, so long as they have not exhausted the allotment of calls afforded them before public health assumes that they are unreachable. For large numbers of people on the call list it may occur that the call list cannot be cleared on a single day. In that case the contact tracing calls will pick up the next day at the point where they left off the day before.

## Experiments

We now present results from several experiments to demonstrate initial validity of SimAEN and that expected trends in output (e.g., total cases) are seen.

### Baseline Validation

The first experiment served to validate our initial settings and to establish the baseline number of cases that are expected from the model. To perform this experiment we set the probability that an individual is running the AEN application (`p_running_app`), the number of contact tracers, (`n_contact_tracers`), and the proportional reduction due to one of the parties, generator or generated, wearing a mask (`mask_effect`) to 0. This combination effectively removes all of the interventions, simulating how people behaved in the early stages of the pandemic. Results are shown in Figure 3.

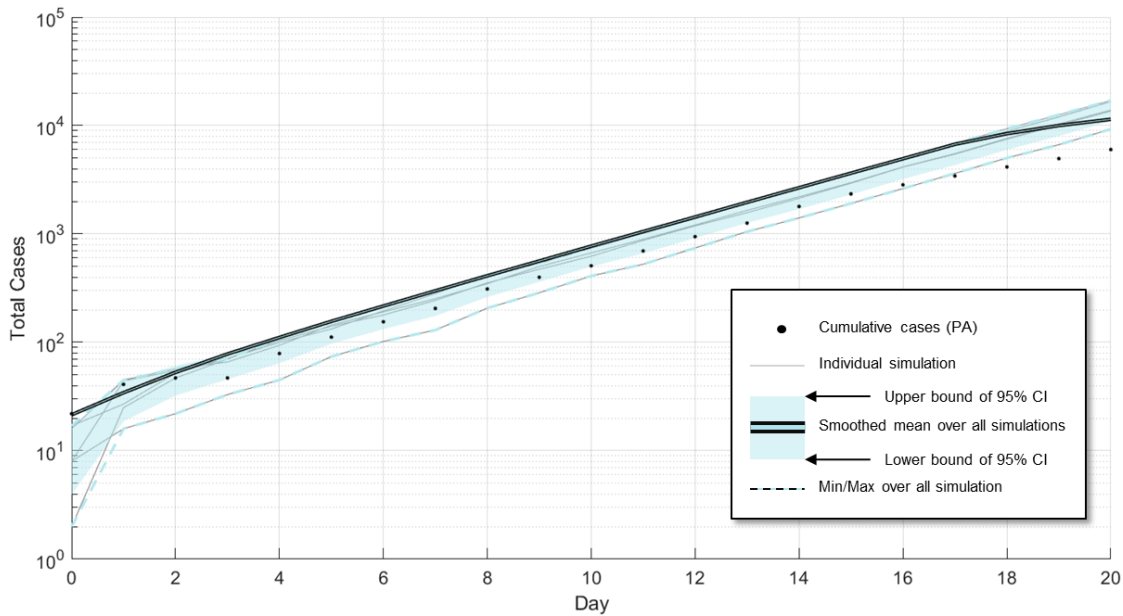

*Figure 3-Plot of the first 20 days of the simulation without targeted interventions along with the Pennsylvania cases for comparison purposes*

The close match seen between the simulation results and the actual number of cases as measured in Pennsylvania, gives us confidence that the model is operating as expected and that the parameter choices are valid.

### Predicting Effect of Masks

Our next experiment involved evaluating the effect of masks on disease progression. To perform this experiment we left the AEN and MCT factors as they were in the baseline experiment above, but set the mask effectiveness (`mask_effect`) to 0.65.

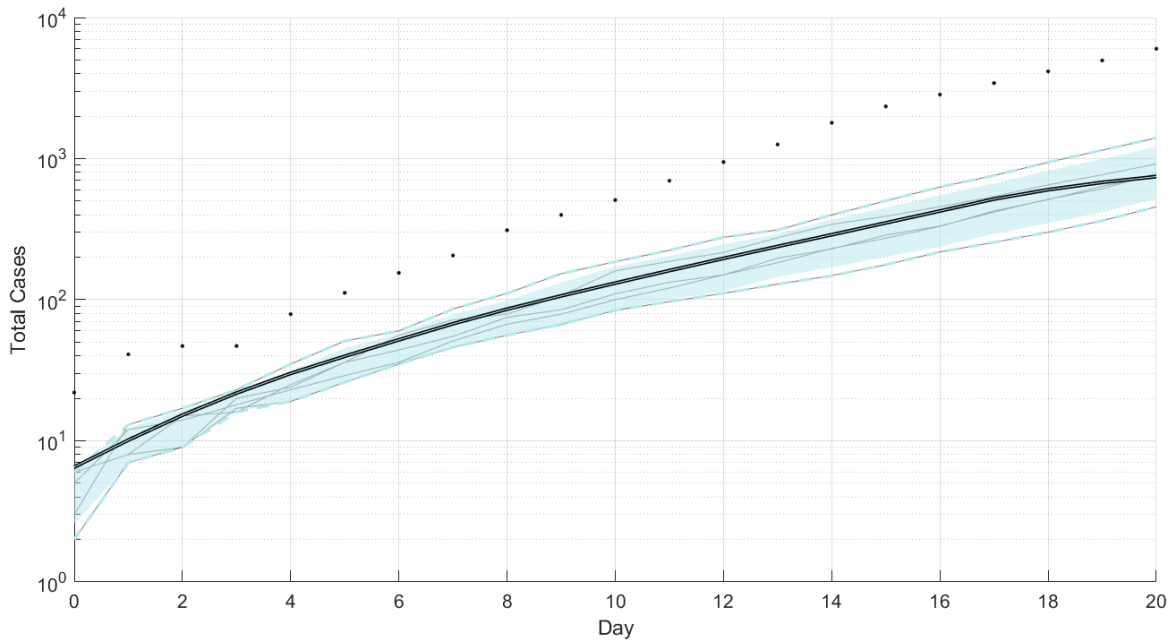

*Figure 4-Plot of 20 days of simulation testing the effect of masks without additional interventions*

As shown in Figure 2, a 0.65 mask effectiveness and a 50% probability of mask adoption in the normal state, we see a 0.5 to 1.0 decade decrease in cases.

As the pandemic has progressed in Pennsylvania, the rate of infection has dropped significantly. To simulate this long term trend we altered the underlying Gaussian distribution from $\mathcal{N}(2,1.1)$ to $\mathcal{N}(0.5,0.25)$. With this lower generation rate we also increased the number of starting cases to 1000 (from 20) to ensure that there are a sufficient number of infected agents for the spread to propagate. This setup with the reduced case rate was used to investigate the long-term effects of AEN, MCT, and their combined operation.

The first such experiment was to turn off MCT and set AEN at 10%. With this low adoption rate we see a decrease of 1670 cases (9%) after 120 days.

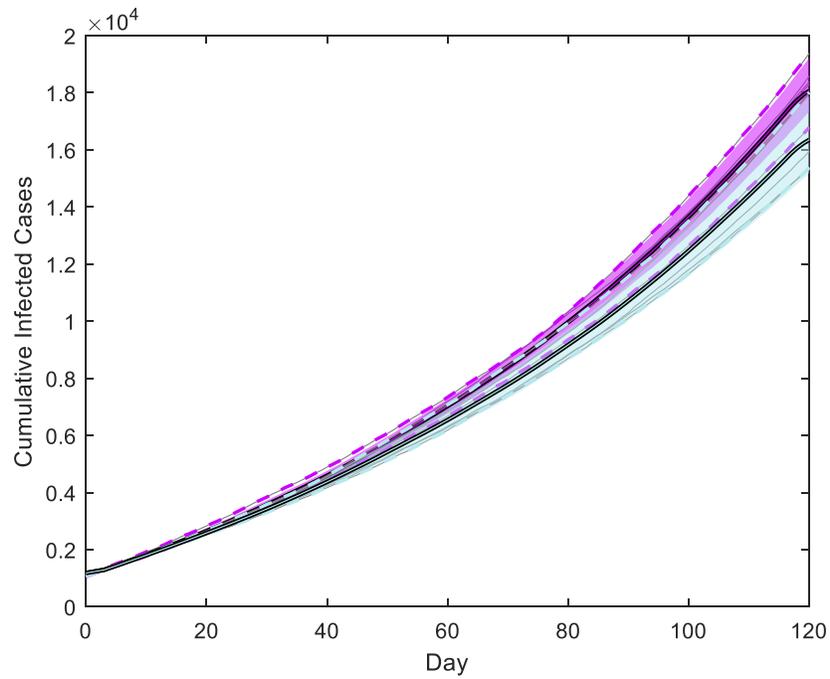

*Figure 5-Baseline (pink) and AEN with 10% adoption rate (blue) cumulative infected cases*

Increasing adoption rate to 50% yields a slightly larger decrease, an average of 2,310 cases (13%).

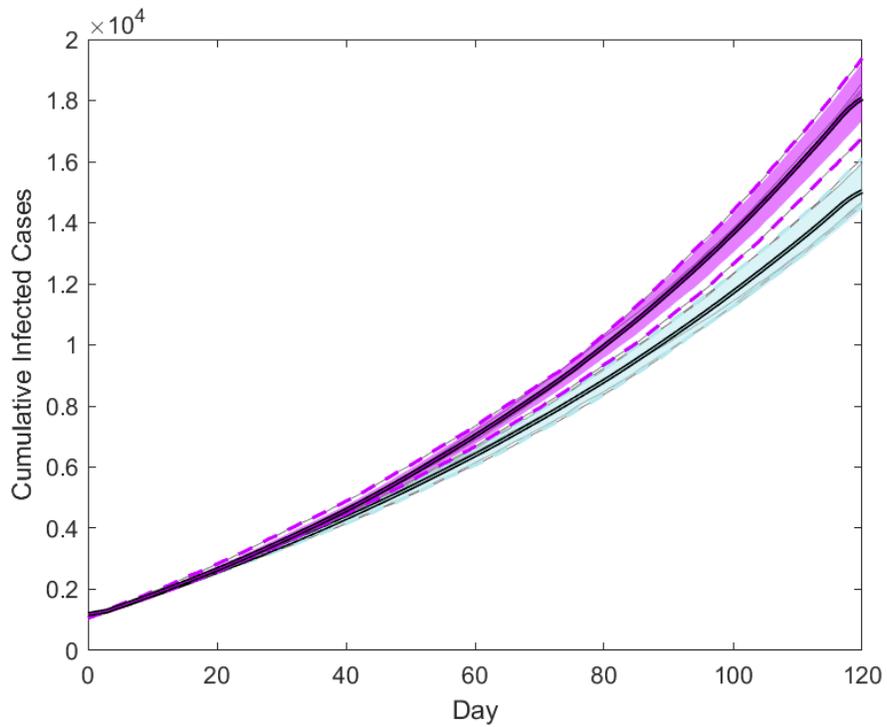

*Figure 6-Baseline (pink) and AEN with 50% adoption rate (blue) cumulative infection cases*

This limited effect is due to several factors. For AEN to affect an individual it requires both the infectious agent and the infected agent to be running the app. The low adoption rate makes this unlikely. This is further compounded by the delay between the generation event and the infected agent receiving a positive test. Significant spread can occur during this time.

To complete evaluation of the individual interventions, we turned on MCT, while leaving AEN and masks off. We chose 1000 contact tracers, which was the estimated number that Pennsylvania will employ. For the relatively small number of cases this is an excessive number of contact tracers and as such a call will be made to every individual on the call list every day.

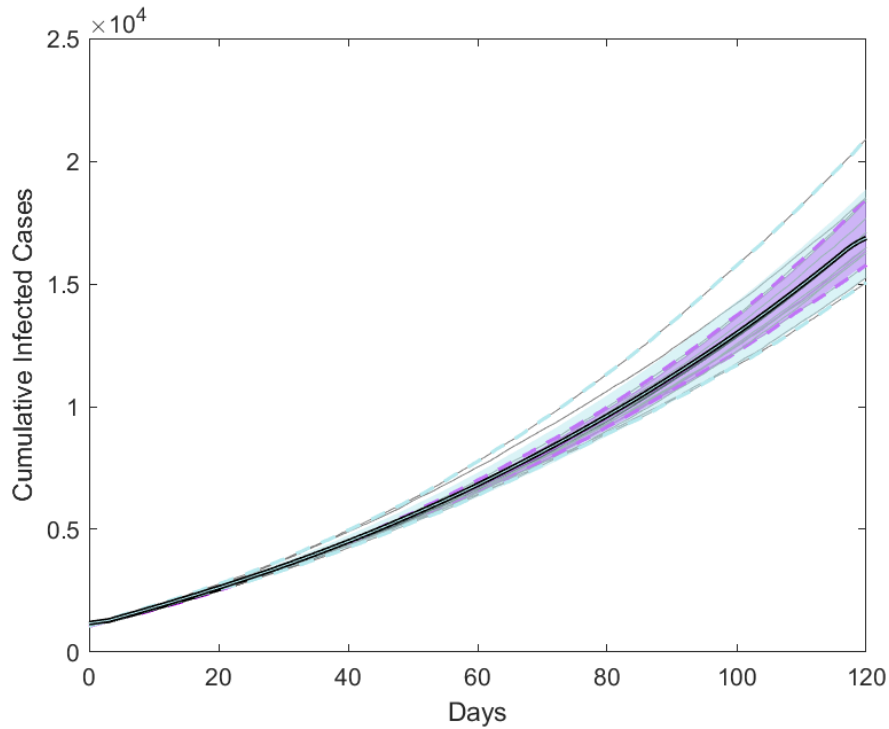

*Figure 7-The effects of MCT (baseline in pink, MCT active in blue). The two curves are essentially identical*

With an only 10% chance of an agent knowing their generator there are simply not enough individuals identified to significantly impact the spread of the disease.

Since AEN prompts agents to get tested, and positive tests drive MCT, we performed an experiment to look at the reciprocal, related effects of these two interventions. For this experiment we set `masks_effect` to 0, but set app adoption to 10% and the number of contact tracers to 1000.

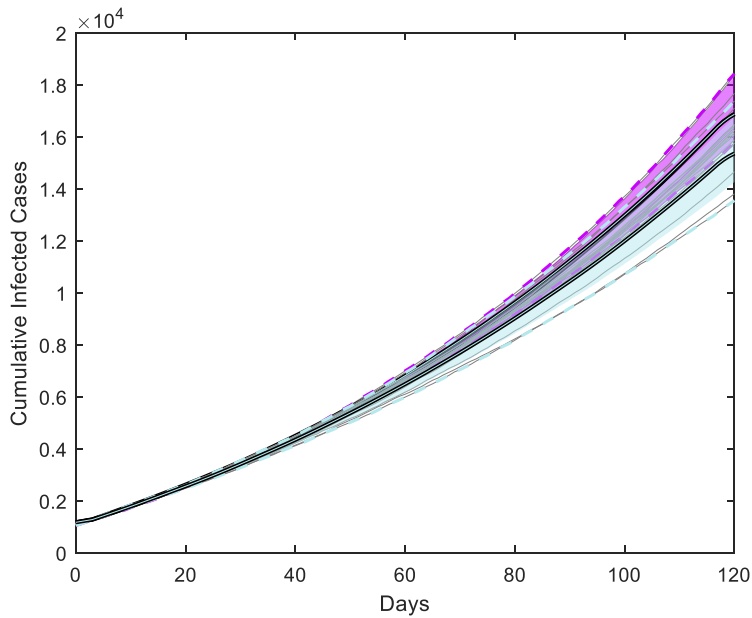

*Figure 8-The effects of combining MCT with AEN at 50% app adoption rate*

We see the effects of the AEN, but there is no noticeable improvement with the MCT despite the interaction between these two interventions.

## Conclusion

We have presented an agent-based model of the effects of AEN and public health on the progression of the COVID-19 disease. As demonstrated in the experiments section, the proposed model produces a good fit for the early pandemic data, as measured in Pennsylvania, indicating that its most critical aspects are a reasonable reflection of reality. Additional simulated effects resulting from varying mask usage, app adoption and initial number of infections match expected behavioral trends and further support confidence in the model.

Some initial conclusions can be drawn from experiments with SimAEN. Because of the widespread asymptomatic spread the most important aspect of any intervention relying on agent behavioral changes is adoption. Widespread masking is the most effective treatment because it reduces the transmission immediately, without the delay associated with AEN and MCT. These delays are a product of testing, but also the time that it takes for symptoms to develop in the generators. During this time there is ample opportunity for a large scale spreading event to occur.

MCT is particularly poor at reducing the number of cases. This is because of the low probability of an infectious agent identifying the agents they infected. Coupling this with the delays permitting transmission events means that the spread cannot be controlled through this method, provided that large scale spread is already occurring.

The best value to be achieved through AEN and MCT may come in the community surveillance that they permit instead of on their direct impact on the effective transmission rate. Public health could use this information to expand community testing efforts, direct messaging, and identify planned potential super-spreader events in these areas.

We anticipate that SimAEN's true value will come from its ability to explore effects of interventions and PH workflow changes in the face of the pandemic. Though initial results from the exploration of disease propagation through SimAEN are promising, we will continue to calibrate model assumptions and validate predictions based on available data.

# Appendix

The probabilities that dictate the generation and lifetimes of individuals are also kept in the World object.

| | |
|---|---|
| The number of cases at the start of the simulation | starting_cases = 20 |
| Number of days that the simulation lasts | end_day = 30 |
| Maximum number of current cases before program stops | max_num_current_cases = 1_500_000 |
| The mean time between an individual being exposed and them becoming infectious | mean_latent_period = 2 (Pierlinck, Linka, Costabal, & Kuhl, 2020) |
| The standard deviation of latent period | sd_latent_period = 0.7 (Pierlinck, Linka, Costabal, & Kuhl, 2020) |
| | |
| The mean time between an individual being exposed and them becoming clinical | mean_incubation_period = 6 (Backer, Klinkenberg, & Wallinga, 2020) |
| The standard deviation of incubation period | sd_incubation_period = 2.3 (Backer, Klinkenberg, & Wallinga, 2020) |
| The probability that a person who has been called by public health will get tested on any given day | p_test_given_call = 0.5 |
| The probability that a person who has no symptoms and has not been notified in any way will get a test | p_test_baseline = 0.1 |
| The probability that a person who has received a notification through the app will get tested on any given day | p_test_given_notification = 0.3 |
| The probability that a person who is symptomatic will get tested on any given day | p_test_given_symptomatic = 0.5 |
| The probability that a person who have been exposed will test positive | p_positive_test_given_exposed = 0.5 |
| The probability that someone who is presymptomatic will test positive | p_positive_test_given_presymptomatic = 0.75 |
| The probability that someone who is symptomatic will test positive | p_positive_test_given_symptomatic = 0.9 (Watson, Whiting, & Brush, 2020) |
| The probability that someone who is asymptomatic will test positive | p_positive_test_given_asymptomatic = 0.9 |
| The probability that a person is running the app | p_running_app = 0.1 |
| The probability that a close contact running the app will successfully handshake with their generator (given the generator is also running the app) | p_app_detects_generator = 0.9 |
| The probability that a call from public health will reach the person | p_successful_call = 0.5 |
| The probability that a person will call public health after a positive test | p_contact_public_health_after_positive_test = 0.5 |
| The probability that a person will call public health after receiving an AEN notification | p_contact_public_health_after_AEN_notification = 0.5 |
| The probability that a person who is running the app who gets a positive test will upload their key to public health | p_upload_key_given_positive_test = 0.75 |
| The probability that an individual will be found using manual contact tracing | p_identify_individual_using_manual_contact_tracing = 0.1 |
| The likelihood an infected person will be asymptomatic | p_asymptomatic_rate = 0.73 (Poletti, et al., 2020) |
| | |
| The average number of contacts that an individual encounters each day.... | |
| ...if they take no precautions - Log-normal distribution | mean_new_cases = 2.0 (Fu, 2005) * |
| | sigma_new_cases = 1.1 (Fu, 2005) * |
| ...if they social distancing / being cautious | mean_new_cases_minimal = 1.1 |
| | sigma_new_cases_minimal = 0.6 |
| ...after entering self-Quarantine | mean_new_cases_moderate = 0.6 |
| | sigma_new_cases_moderate = 0.3 |
| ...after entering self-isolation | mean_new_cases_maximal = 0.3 |
| | sigma_new_cases_maximal = 0.1 |
| | |
| | 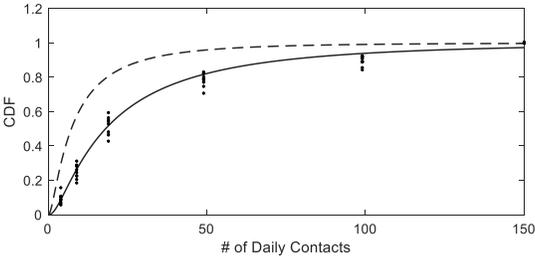
Curve fit (full line) of data from Fu, 2005 (dots) has $\mu = 2.9$ and $\sigma = 1.1$. Our operating point (dashed line) was selected based on fit to Pennsylvania cases data. The lower mean is attributable to several conditions:
- The counts from Fu, 2005 are not unique interactions. Since most people interact with their family and coworkers regularly we would anticipate a lower mean number of unique contacts
- The Pennsylvania case counts are for the entire state which includes many rural areas which may have significantly lower numbers of interactions than those found in the more urban study
- Due to the high number of asymptomatic cases, the number of reported cases is likely much lower than the true number of cases
- The first cases in Pennsylvania (March 10, 2020) were several months after reports from Wuhan (January 2020), which may have prompted changes in behavior of the population

It is our belief that the combination of these factors justify the lower mean. |
| The probability that a true contact event involving an infected person will result in infection | p_transmission_given_no_masks = 0.07 (Laxminarayan, et al., 2020) |
| | |
| The False Discovery Rate (FDR), used to create additional false positives picked up automatically by the system. 0.5 will equal new cases. | false_discovery_rate = 0.25 |
| | |
| Probabilities associated with entering various levels of restricted movement given the person is symptomatic | p_maximal_restriction_given_symptomatic = 0.5 |
| | p_moderate_restriction_given_symptomatic = 0.75 |
| | p_minimal_restriction_given_symptomatic = 0.9 |
| | |
| Probabilities associated with entering various levels of restricted movement given the person receives a positive test | p_maximal_restriction_given_positive_test = 0.75 |
| | p_moderate_restriction_given_positive_test = 0.85 |
| | p_minimal_restriction_given_positive_test = 0.95 |
| | |
| Probabilities associated with entering various levels of restricted movement given the person is successfully called by PH | p_maximal_restriction_given_PH_call = 0.5 |
| | p_moderate_restriction_given_PH_call = 0.75 |
| | p_minimal_restriction_given_PH_call = 0.9 |
| | |

| Probabilities associated with entering various levels of restricted movement given the person is notified by AEN | p_maximal_restriction_given_AEN_notification = 0.5 |
|---|---|
| | p_moderate_restriction_given_AEN_notification = 0.75 |
| | p_minimal_restriction_given_AEN_notification = 0.9 |
| | |
| The mean and standard deviation of number of days that it takes for a test to get back (normal distribution) | test_delay = 4 (Lazer, et al., 2020) |
| | test_delay_sigma = 1 |
| | |
| Number of days it takes to be sure of recovery from infection | recovery_length = 17 (Pierlinck, Linka, Costabal, & Kuhl, 2020) |
| The number of contact tracers | n_contact_tracers = 1000 (Pennsylvania Pressroom, 2020) |
| The number of hours each contact tracer can spend on calling in a day | work_day_length = 4 |
| The number of time PH will try to contact an individual before giving up | max_call_attempts = 3 |
| The length of time that a missed call takes (in hours) | missed_call_time = 0.05 |
| The length of time that a contact tracer takes to perform contact tracing on an index case (in hours) | index_trace_call_time = 1 |
| The length of time that a missed call takes (in hours) | alert_call_time = 0.1 |

The World object is comprised of the following variables.

| population | An array of all of the currently active individuals |
|---|---|
| index_cases | List of cases that have been tested positive and submitted their key via the app |
| test_queue | List of people waiting for a test |
| call_list | The list of individuals that have had a contact event, are running the app, are known about, and need to be called by public health |
| total_available_call_time | The total amount of man-hours that are worked by contact tracers each day |

The Individual object has these variables set.

| myWorld | The world that the agent belongs to |
|---|---|
| generator | The individual that transmitted the disease to this individual |
| daysInSystem | The number of days that an individual has existed |
| index | A unique identifier of the object |
| test_countdown | The number of days before a test will come back |
| dayTested | The date that the person was tested |
| hasApp | Whether the individual is running the AEN application |
| appDetectsGenerator | Whether the person who transmitted the disease to this individual was detected |
| appDetectedByGenerator | Whether the person who transmitted the disease to this individual detected the infected individual |
| notified | If the individual has received an alert from AEN |

| | | | |
|---|---|---|---|
| called | If the individual has been called by public health | | |
| hasBeenManualContactTraced | If the person has had manual contract tracing performed | | |
| tested | If the person has been tested | | |
| statusAtTest | The infection status of the person on the day they were tested | | |
| testResultReceived | If the person has gotten their test results back | | |
| keyUploaded | If the person decided to upload their key to the AEN network after a positive test | | |
| testResult | The result that the individual received once their test returned | | |
| infectionStatus | The current disease state of the individual | | |
| | | Uninfected | The individual did not contract the disease |
| | | Exposed | The individual has been infected |
| | | Presymptomatic | The individual has become infectious but is not showing symptoms |
| | | Asymptomatic | The individual is infectious but does not show symptoms |
| | | Symptomatic | The individual is infectious and shows symptoms |
| | | Recoverd | The individual has recovered from the disease |
| behavior | How the individual is interacting with the world | | |
| | | Nominal | |
| | | Minimal Restriction | |
| | | Moderare Restriction | |
| | | Maximal Restriction | |
| callAttempts | The number of times that public health has tried to call this individual | | |
| daysOnCallList | The number of days that the individual has been on the call list | | |
| latentPeriod | The length of time that the individual will be in the Exposed state | | |
| incubationPeriod | The length of time that the individual will be in the Presymptomatic | | |